\documentclass[prb,twocolumn,showpacs,bm,floatfix]{revtex4}
\usepackage{epsfig}
\begin{document}
\title{Phase separation in the combined Falicov-Kimball and static 
Holstein model }
\date{\today}
\author{B.~M.~Letfulov}
\affiliation{Institute of Metal Physics, Kovalevskaya Str. 18, Yekaterinburg,
620219, Russia}
\author{J.~K.~Freericks}
\email{freericks@physics.georgetown.edu}
\homepage{http://www.physics.georgetown.edu/~jkf}
\affiliation{Department of Physics, Georgetown University, Washington, DC
20057}
\begin{abstract}
The Falicov-Kimball model has been known to phase separate when the
correlation strength is large and the temperature is low.  We examine
the stability of phase separation under the influence of an additional
electron-phonon interaction that tries to force the electron density
to be homogeneous.  We find that the electron-phonon interaction
eventually destroys the phase-separated state once it becomes larger
than some critical value.  The results of this work may have an influence
on theories of collossal magentoresistance materials.
\end{abstract}
\maketitle

The phenomenon of strong electron correlations driving phase separation 
has been long conjectured in the Falicov-Kimball 
model\cite{freericks_falicov_1990} and in the Hubbard 
model\cite{emery_kivelson_lin_1990}.  Little progress has been made on
rigorously establishing phase separation in the Hubbard model, but
it has been proved to exist in the Falicov-Kimball model in
one dimension\cite{lemberger_1992}, in infinite 
dimensions\cite{freericks_gruber_macris_1999}, and in all other 
dimensions\cite{freericks_lieb_ueltschi_2002a,freericks_lieb_ueltschi_2002b}.
In spite of these rigorous results, there has still been only limited
calculations of phase diagrams and the transition temperature as a function
of the parameters in the model.  Essentially all finite-temperature calculations
have been performed in the infinite-dimensional 
limit\cite{brandt_mielsch_1991,letfulov_1999,freericks_gruber_macris_1999,freericks_lemanski_2000}.
Here we continue with this tradition, and illustrate how competing (finite)
interactions can destroy the phase-separated state, even if the 
Falicov-Kimball interaction strength is infinitely large.

The system we choose to study is the harmonic version of the spinless
Falicov-Kimball--static-Holstein model\cite{falicov_kimball_1969,holstein_1959}.
The Hamiltonian is
\begin{eqnarray}
H&=&-\frac{t^*}{2\sqrt{d}}\sum_{\langle i,j\rangle}c^\dagger_ic_j+(E_f-\mu)
\sum_i f^\dagger_if_i+U\sum_ic^\dagger_ic_if^\dagger_if_i\cr
&+&\frac{1}{2}\kappa\sum_ix_i^2+\sum_i(gx_i-\mu)(c^\dagger_ic_i-\rho_e),
\label{eq: ham}
\end{eqnarray}
where $c^\dagger_i$ ($c_i$) is the creation (annihilation) operator for
an itinerant electron at site i, $f^\dagger_i$ ($f_i$) is the creation
(annihilation) operator for a localized electron at site i,
$t^*/2\sqrt{d}$ is the hopping integral between nearest-neighbor sites
(each pair is summed over twice for hermiticity), $E_f$ is the local
f-electron site energy, $U$ is the Falicov-Kimball interaction, $x_i$ is
the static phonon coordinate, $\kappa$ is the spring constant (chosen to be 1 
here), and
$g$ is the electron-phonon interaction.  The symbol $\rho_e$ denotes the
average 
itinerant-electron concentration $\langle c^\dagger c\rangle$. We examine the
system in the infinite-dimensional limit\cite{metzner_vollhardt_1989} and
take $t^*$ as our energy unit. A common chemical potential $\mu$ is used to
fix the total electron concentration (itinerant plus localized). Note that
we have chosen the phonons to interact with the local charge fluctuation,
which is a common choice, but not the only one that could be made.  Since
the term that is subtracted depends on the localized electron concentration
it is not just a constant that can be included or removed; it has a 
nontrivial effect on the value of the Helmholz free energy, and thereby on
the phase separation.  For concreteness we solve our system of
equations on an infinite-coordination Bethe lattice, which has a
semicircular electronic density of states 
\begin{equation}
\rho(\epsilon)=\frac{1}{2\pi}\sqrt{4-\epsilon^2}.
\label{eq: dos}
\end{equation}

The system is solved by employing dynamical mean-field theory.  The procedure
simply combines the strategy used for the Falicov-Kimball 
model\cite{brandt_mielsch_1989} with that used for the static Holstein 
model\cite{millis_littlewood_shraiman_1995,millis_mueller_shraiman_1996}.
We summarize the procedure here to standardize our notation.  Our interest
is in the $U\rightarrow\infty$ limit.  In that case, the itinerant electrons
and the localized electrons are not allowed to sit at the same lattice
site.  The dynamical mean-field theory is straightforward to evaluate.
The partition function becomes
\begin{eqnarray}
Z&=&\int_{-\infty}^{\infty}dx2e^{-\frac{\beta}{2}(gx-\mu)}
\prod_n\frac{G_0^{-1}(i\omega_n)-gx}{i\omega_n}\cr
&\times&\exp\{-\beta[-(gx-\mu)\rho_e+\frac{1}{2}\kappa x^2]\}\cr
&+&\int_{-\infty}^{\infty}dx2e^{-\frac{\beta}{2}(gx-\mu+U)}
\prod_n\frac{G_0^{-1}(i\omega_n)-gx-U}{i\omega_n}\cr
&\times&\exp\{-\beta[-(gx-\mu)\rho_e+E_f-\mu+\frac{1}{2}\kappa x^2]\},
\label{eq: partition}
\end{eqnarray}
with $\beta=1/T$ the inverse temperature, 
$G_0(i\omega_n)=1/(i\omega_n+\mu-\lambda_n)$ the effective medium,
$i\omega_n=i\pi T(2n+1)$ the Fermionic Matsubara frequency, and
$\lambda(i\omega_n)=\lambda_n$ the dynamical mean field.  Once the
partition function is determined (in the limit $U\rightarrow\infty$), then one can 
calculate the single-particle electronic Green's function via
\begin{equation}
G(i\omega_n)=G_n=\int_{-\infty}^{\infty}dx\frac{w_0(x)}{G_0^{-1}(i\omega_n)-gx},
\label{eq: green1}
\end{equation}
with $w_0(x)$ the phonon distribution function (when no localized-electron occupies
the lattice site), which satisfies
\begin{eqnarray}
w_0(x)&=&2e^{-\frac{\beta}{2}(gx-\mu)}
\prod_n\frac{G_0^{-1}(i\omega_n)-gx}{i\omega_n}\cr
&\times& \exp\{-\beta[-(gx-\mu)\rho_e+\frac{1}{2}\kappa x^2]\}/Z;
\label{eq: w0}
\end{eqnarray}
the integral of $w_0(x)$ satisfies a sum rule
\begin{equation}
1-\rho_f=\int_{-\infty}^{\infty}dxw_0(x).
\label{eq: sumrule}
\end{equation}
[The sum rule is not one, because we have taken the limit $U\rightarrow\infty$.
If $U$ is finite, we would have a second phonon distribution function $w_1(x)$
to consider when calculating the Green's function, which enters when the site is 
occupied by a localized-electron. In the $U=\infty$ limit, we can always rewrite
$w_0(x)=P(x)-w_1(x)$ where the integral of $P(x)$ is one and can be interpreted
as the $U=\infty$ phonon distribution function, and $w_1(x)$ integrates to
the localized electron concentration.  But in this work, we never distinguish these
objects, so there is no need to elaborate further on them.]
To complete the loop for the self-consistent solution of the problem,
we need to note that Dyson's equation is
\begin{equation}
G_0^{-1}(i\omega_n)=G_n^{-1}+\Sigma_n,
\label{eq: dyson}
\end{equation}
and
\begin{equation}
G_n=\int_{-2}^{2}d\epsilon \rho(\epsilon)
\frac{1}{i\omega_n+\mu-\Sigma_n-\epsilon},
\label{eq: glocal}
\end{equation}
with $\Sigma(i\omega_n)=\Sigma_n$ the itinerant-electron self energy.
This provides a closed set of equations that is needed to solve the
many-body problem.

The computational algorithm that we follow is the iterative procedure
introduced by Jarrell\cite{jarrell_1992}: (i) we begin with the 
self energy set equal to zero; (ii) Eq.~(\ref{eq: glocal}) is used
to determine $G_n$; (iii) Eq.~(\ref{eq: dyson}) is used to extract the
effective medium $G_0$; (iv) Eqs.~(\ref{eq: partition}), (\ref{eq: w0}),
and (\ref{eq: green1}), then determine $G_n$; and (v) Eq.~(\ref{eq: dyson})
is employed to extract the new self energy.  Then steps  (ii-v) are
repeated until convergence is reached.  Once the equations have
converged on the imaginary axis, one can take the function $w_0(x)$
and the analytic continuation of Eqs.~(\ref{eq: glocal}), (\ref{eq: dyson}),
and (\ref{eq: green1}) to determine the Green's function on the real
axis as well.  

In order to study phase separation, we also need to calculate the free energy.
This is accomplished by following the procedure of Brandt and 
Mielsch\cite{brandt_mielsch_1991}.  The Helmholz free energy
satisfies
\begin{eqnarray}
F&=&-T\Biggr [ \ln Z+\sum_n \ln G_n\cr
&-&\sum_n\int_{-2}^2d\epsilon \rho(\epsilon)
\ln\frac{1}{i\omega_n+\mu-\Sigma_n-\epsilon}\Biggr ]\cr
&-&(E_f-\mu)\rho_f.
\label{eq: free}
\end{eqnarray}
It is important to include a sufficient number of Matsubara frequencies in the
summation to achieve convergence of the summation; indeed this ultimately limits
the accuracy of our results.

It is well known that if both $\rho_e$ and $\rho_f$ lie between zero and one,
then the ground state is phase separated\cite{freericks_gruber_macris_1999}.
The key point in this work is that we fix $E_f$.  If the chemical potential
lies below $E_f$ then the ground state contains only itinerant electrons and
is not phase separated.  As the chemical potential is increased, we reach
a critical value of $\mu$ where the average localized electron filling becomes
nonzero, and the system goes into a phase-separated state.  It is straightforward
to determine the value of this critical chemical potential when $g=0$.  The 
interacting density of states is just the semicircle in Eq.~(\ref{eq: dos}) when 
there are no localized electrons and it becomes
\begin{equation}
\rho_{int}(\epsilon)=\frac{1}{2\pi}\sqrt{4(1-\rho_f)-\epsilon^2}
\label{eq: dos2}
\end{equation}
when there are $\rho_f$ interacting electrons. So we must compare the energy
with no localized electrons $E_{gs}=-(4-\mu^2)^{3/2}/6\pi$ to the mixture of states
with all localized particles and no itinerant particles (with weight $\rho_f$)
and the state with no localized particles but with an electron filling of
$\rho_e=\rho_{tot}-(1-\rho_{tot})\rho_f$ (with weight $1-\rho_f$) 
in the limit where $\rho_f$ approaches zero.  Note that one must adjust
the chemical potential $\mu^*$ of the mixed state, to have the correct electron
density.
After performing some straightforward algebra, we find a transcendental equation
for the critical value of the chemical potential
\begin{eqnarray}
E_f&=&-\frac{(4-\mu^2)^{\frac{3}{2}}}{6\pi}-\mu \Biggr ( -\frac{1}{2}+\frac{1}{\pi}
\sin^{-1}\frac{\mu}{2}\cr
&+&\frac{\mu}{4\pi}\sqrt{4-\mu^2}\Biggr ).
\label{eq: transcend}
\end{eqnarray}
As $E_f$ ranges from $-2$ to 0, the critical total electron density for phase
separation ranges from 0 to 1; for $E_f>0$ the system does not phase separate
as $T\rightarrow 0$.

\begin{figure}[t]
\epsfxsize=3.0in
\epsffile{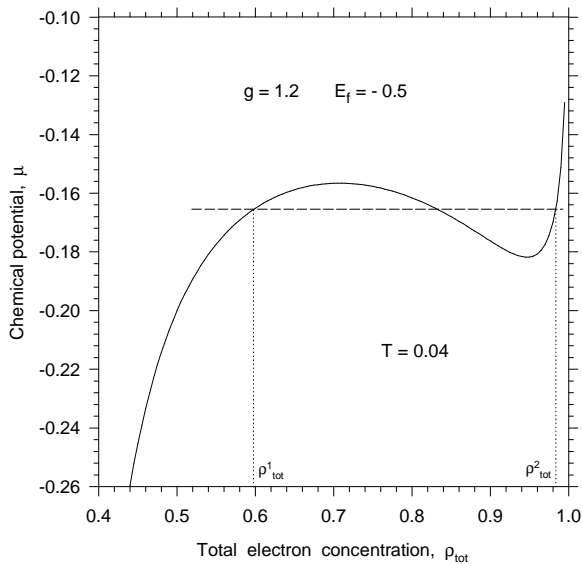}
\caption{
\label{fig: 1}The chemical potential is plotted versus the total electron
filling when $g=1.2$,
$E_f=-0.5$ and $T=0.4$.  Note the double-valued chemical potential 
indicating phase separation. The Maxwell construction is also shown
in the figure by the dashed line.  The total area of the solid curve above and
below the dashed line is equal to zero. The dotted lines indicate the
total electron densities of the states that form the mixture.}
\end{figure}

We will examine the case with $E_f=-0.5$ here.
Eq.~(\ref{eq: transcend}) is solved by $\mu=-0.145$ and
$\rho_e+\rho_f=0.454$ for the lower boundary where phase separation begins.
Hence phase separation enters when $\rho_{tot}>0.454$.
To determine the phase separation, we must perform a Maxwell construction.  If we
plot the electron filling as a function of the chemical potential, we will find
a multivalued function in the region of phase separation.  One way to perform
the Maxwell construction
is by drawing a line that has equal areas above and below, with the points
of intersection corresponding to the densities of the phases that compose the 
mixture of the phase-separated state (the other way is to construct the convex
hull of the free-energy curve; both methods yield the same results).  The 
double-valued function and the Maxwell construction are plotted for 
$g=1.2$ and $T=0.04$ in Fig.~\ref{fig: 1}.  
It is nontrivial to generate such a curve.
The technical difficulty arises from the fact that for a fixed value of $E_f$,
there are multiple solutions of $\rho_f$ which have the same $\rho_{tot}$
when one is in the phase separated region.  Hence, we sometimes need to either
adjust $\rho_f$ and calculate $E_f$ once the equations have converged, or
employ a two-dimensional rootfinder for $\mu$ and $\rho_f$ to get the equations
to converge.  Even so, we find regions of parameter space where we are unable to 
stabilize the algorithm.

\begin{figure}[t]
\epsfxsize=3.0in
\epsffile{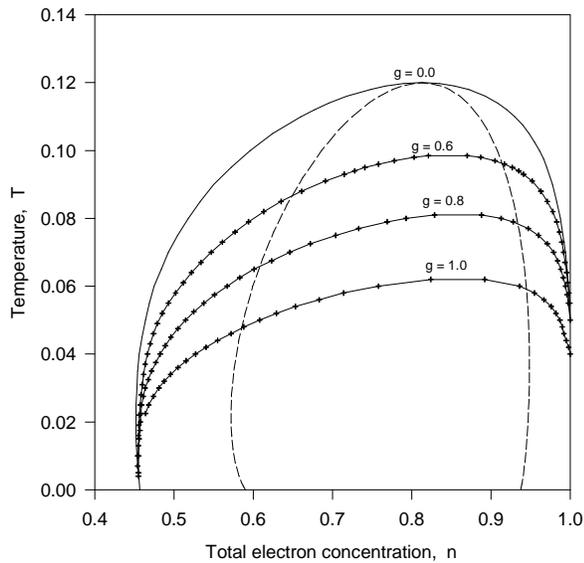}
\caption{
\label{fig: 2} The first-order phase transition as a function of the total
electron filling (the solid dots show the results of the Maxwell construction).  
We examine four cases $g=0$, $g=0.6$, $g=0.8$, and $g=1.0$.
The dashed line is the spinodal decomposition temperature (found by calculating
the divergence of the $k=0$ charge susceptibility) for the $g=0$ case.}
\end{figure}

We examine the transition temperature for different electron-phonon
coupling strengths in Fig.~\ref{fig: 2}.  
The phase-separated region hardly changes at 
zero temperature as the electron-phonon coupling increases, but the transition
temperatures are sharply reduced.  We run into numerical problems with the
stability of our iterative solutions when $g$ is increased to 1.2 and above.
We are unable to find any strategy to solve for the phase separation at low
temperature in this region of parameter space.  We expect, however, that the
phase separation continues to be suppressed, and conjecture that there is a
critical value of the electron-phonon coupling where the phase separation
disappears, but we cannot determine it's value with our numerical algorithms.
In addition to the binodal (first-order) phase transition temperature, we also
include the spinodal decomposition temperature (below which the system cannot
be supercooled) for the $g=0$ case.  This is found by determining where the
$k=0$ charge susceptibility diverges.   Of course the binodal and spinodal
temperatures must agree at the peak of the curve.

The disappearance of the phase separation occurs when
the electron-phonon interaction energy becomes similar in magnitude
to the kinetic-energy gain of the phase separated state.  This occurs at a 
finite value of $g$ because the kinetic energy gain is finite, and does not
diverge as $U\rightarrow\infty$.

To understand this phenomenon further, we examine in detail the system
in the low-temperature limit.  We already saw that when $g=0$ there is a minimum
total electron density that must be reached before phase separation
occurs.  Similarly, when $g\ne 0$, we find that such a critical density continues
to hold for small $g$.  This occurs because, as $T$ is lowered, the function
$w_0(x)$ becomes sharply peaked at $x=0$, eventually becoming a delta function
at $T=0$.  Hence, the system remains a Fermi gas in this regime of parameter
space.  As $g$ is increased to a critical value, $w_0(x)$ develops a local
minimum at $x=0$ and two maxima, one at $x<0$ and one at $x>0$.  At this point,
the ground state becomes a non-Fermi-liquid, and we believe
that the non-Fermi-liquid character
suppresses the phase separation, ultimately eliminating it.  To be more
quantitative, we compute
\begin{equation}
T\frac{d\ln w_0(x)}{dx}=g(\rho_e-\frac{1}{2})-\kappa x-gT\sum_n
\frac{1}{G_0^{-1}(i\omega_n)-gx}.
\label{eq: lnw}
\end{equation}
When this derivative vanishes, we have an extremum of the phonon distribution
function.  When $\rho_f=0$, this equation always has a solution at $x=0$
because the sum over Matsubara frequencies equals $\rho_e-1/2$.  If $g$ is
increased, we reach a point where the second derivative becomes positive, and
the $x=0$ solution becomes a local minimum instead of a maximum.  At this point
two nonzero $x$ solutions exist, and the system becomes a non-Fermi-liquid.
The second derivative (at $x=0$) is
\begin{equation}
T\frac{d^2\ln w_0(x=0)}{dx^2}=-\kappa-g^2T\sum_n G_0^{2}(i\omega_n),
\label{eq: lnw2}
\end{equation}
which becomes 
\begin{eqnarray}
T\frac{d^2\ln w_0(x=0)}{dx^2}&=&-\kappa-g^2\int_{-2}^{\mu}d\epsilon\rho(\epsilon)
{\rm Re}G(\epsilon)\cr
&=&-\kappa+g^2\frac{(4-\mu^2)^{\frac{3}{2}}}{12\pi}
\label{eq: lnw3}
\end{eqnarray}
in the low-temperature limit.  For $\mu=-0.145$ the critical value of
$g$ to enter the non-Fermi-liquid state is 2.18.

It is useful to discuss the impact of this work on models for the
collossal magnetoresistance manganite materials. There are a number
of different elements that appear to be important in describing the CMR
materials.  All models start with double-exchange\cite{furukawa_1994}, 
but then need something
else to enhance the metal-insulator transition.  Three main ideas have
emerged to do this: a strong electron-phonon 
interaction\cite{millis_littlewood_shraiman_1995,millis_mueller_shraiman_1996,edwards_2002}; 
strong disorder scattering\cite{allub_alascio_1996,varma_1996,allub_alascio_1997}; 
and proximity to phase separation\cite{dagotto_2001}.  The results of this 
investigation
indicate that the three elements are not fully independent.  Indeed, strong
disorder scattering can lead to phase separation, and strong electron-phonon
scattering can suppress the phase separation.  One of the consequences of this
work is that one needs to carefully investigate the implications of all three
of these physical effects, and how they interrelate, to construct a meaningful
theory for the manganites.

In conclusion, we have analyzed the suppression of the phase separation transition
in a combined Falicov-Kimball--static-Holstein model.  We see that increasing the
electron-phonon coupling forces the system to be more homogeneous and suppresses
the phase separation.  We conjecture that a finite value of $g$ will be sufficient
to remove phase separation entirely, but are unable to access that region of the
phase diagram due to numerical difficulties.  Our results could have an impact
on CMR materials that have both disorder and electron-phonon interactions.
Any phase separation induced by the disorder may be suppressed by the 
electron-phonon coupling.

\textit{Acknowledgments:} B.M.L. acknowledges support from the Project for
Supporting of Scientific Schools, N 00-15-96544.
J.K.F. acknowledges support from the National Science
Foundation under Grant No. DMR-9973225. 

\bibliography{fk_dmft.bib}

\end{document}